\documentclass[12pt]{article}

\usepackage{epsf}
\usepackage[dvips]{graphicx}
\usepackage{dcolumn}
\newcommand{\Det}{{\rm Det}}
\newcommand{\Tr}{{\rm Tr}}
\newcommand{\tr}{{\rm tr}}

\newcommand{\be}{\begin{equation}}
\newcommand{\ee}{\end{equation}}
\newcommand{\bea}{\begin{eqnarray}}
\newcommand{\eea}{\end{eqnarray}}
\newcommand{\ba}{\begin{array}{l}}
\newcommand{\ea}{\end{array}}

\newcommand{\lab}[1]{\label{#1}}
\newcommand{\re}[1]{(\ref{#1})}

\begin{document}

\begin{center}
{\bfseries INSTANTON VACUUM BEYOND CHIRAL LIMIT}\footnote{Talk given 
at the XVII International Baldin Seminar
"Relativistic Nuclear Physics and Quantum Chromodynamics", 
Sept.27-Oct.2, 2004 (JINR, Dubna, Russia). 
}

\vskip 5mm

M. Musakhanov
\vskip 5mm

{\small
 {\it
 Department of Physics,
Pusan National University, \\
609-735 Pusan,
Republic of Korea\\
\& Theoretical  Physics Dept,
Uzbekistan National University,\\
 Tashkent 700174, Uzbekistan
}
\\
{\it
E-mail: musakhanov@pusan.ac.kr
}}
\end{center}

\vskip 5mm

\begin{center}
\begin{minipage}{150mm}
\centerline{\bf Abstract}
In this talk it is discussed the derivation of low--frequencies part
of quark determinant and partition function. As a first application,
quark condensate is calculated beyond chiral
limit with the account of $O(m)$, $O(\frac{1}{N_c})$, $O(\frac{1}{N_c}m)$ and
$O(\frac{1}{N_c}m\ln m)$ corrections. It was demonstrated
complete correspondence of the results to chiral perturbation
theory.
\end{minipage}
\end{center}

\vskip 10mm

\section*{Introduction}
Instanton vacuum model assume that QCD vacuum is filled not only
by perturbative but also very strong non-perturbative fluctuations
-- instantons. This model provides a natural mechanism for the
spontaneous breaking of chiral symmetry (SBCS) due to the
delocalization of single-instanton quark zero modes in the
instanton medium.  The model   is described by two main parameters --
the average instanton size $\rho\sim 0.3\, fm$ and average
inter-instanton distance $R\sim 1\, fm$. These values was found
phenomenologically \cite{Shuryak:1981ff} and theoretically
\cite{Diakonov:1983hh} and was confirmed by lattice measurements
\cite{Chu:vi,Negele:1998ev,DeGrand:2001tm,Faccioli:2003qz,Bowman:2004xi}.
On the base of this model was developed effective action approach
\cite{Diakonov:1985eg,Diakonov:1995qy,Diakonov:2002fq}, providing
reliable method of the calculations of the observables in hadron
physics at least in chiral limit.

On the other hand, chiral perturbation theory makes a theoretical
framework incorporating the constraints on low-energy behavior of
various observables based on the general principles of chiral
symmetry and quantum field theory \cite{Gasser:1983yg}.

It is natural expect, that  instanton vacuum model leads to  the
results compatible with chiral perturbation theory.

One of the most important quantities related with SBCS is the
vacuum quark condensate $<\bar qq>$, playing also important
phenomenological role in various applications of QCD sum rule
approach. Previous investigations \cite{Novikov:xj} shows that
beyond chiral limit and at small current quark mass $m\sim few \,
MeV$ these quantity receive large so called chiral log
contribution $\sim \frac{1}{N_c}\,m\,\ln\,m$ with fixed model
independent coefficient. On the typical scale $1\,GeV$ it become
leading correction since $|\frac{1}{N_c}\ln\,m|\geq 1$. It was
shown, that this correction is due to pion loop contribution
\cite{Novikov:xj,Gasser:1983yg}.

So, to be consistent we have to calculate simultaneously all of
the corrections of order $m,\,\,\frac{1}{N_c},\,\,
\frac{1}{N_c}\ln\,m $ in order to find quark condensate beyond
chiral limit.

In our previous papers
\cite{Musakhanov:1998wp,Musakhanov:vu} on the
base of low--frequencies part of light quark determinant
$\Det_{low},$  obtained in
\cite{Lee:sm,Diakonov:1985eg,Schafer:1996wv}, was derived
effective action. In this framework was investigated current
quark mass $m$ dependence of the  quark condensate, but without
meson loop contribution \cite{Musakhanov:vu}.

In the present work we refine the derivation of the
low--frequencies part of light quark determinant $\Det_{low}$.
The following averaging  of $\Det_{low}$ over instanton
collective coordinates is done independently over each instanton
thanks to small packing parameter $\pi(\frac{\rho}{R})^4\sim 0.1$
and also by introducing constituent quarks degree of freedoms
$\psi .$ This procedure leads to the light quarks partition
function $Z[m].$ We apply bosonisation procedure to $Z[m]$, which
is exact one for our case $N_f=2$ and  calculate partition
function $Z[m]$ with account of meson loops.  This one provide us
the quark condensate with desired $O(m),\,\,O(\frac{1}{N_c}),\,\,
O(\frac{1}{N_c}m\,\ln\,m )$  corrections.

\section*{Low--frequencies part of light quark determinant}
The main assumption of previous works
\cite{Diakonov:1985eg,Diakonov:1995qy,Diakonov:2002fq} (see also
review \cite{Schafer:1996wv}) was that at very small $m$ the quark
propagator in the single instanton field $A_i$ can be approximated
as:
\begin{equation}
S_I (m\sim 0) \approx \frac{1}{i\hat\partial} +
\frac{|\Phi_{0I}><\Phi_{0I}|}{im}
\end{equation}
It gives proper value for the $ <\Phi_{0I}|S_I (m\sim
0)|\Phi_{0I}> =\frac{1}{im},$ but in $ S_I (m\sim 0)|\Phi_{0I}> =
\frac{|\Phi_{0I}>}{im} +\frac{1}{i\hat\partial}|\Phi_{0I}> $
second extra term has a wrong chiral properties. We may neglect
by this one only for the $m\sim 0 .$

 At the present case of non-small $m$ we assume:
\begin{equation}
S_I\approx S_{0} + S_{0}i\hat
\partial \frac{|\Phi_{0I}><\Phi_{0I}|}{c_I} i\hat
\partial S_{0},\,\,\, S_0 = \frac{1}{i\hat\partial +im}
\lab{Si}
\end{equation}
where
\begin{equation}
c_I=-<\Phi_{0I}|i\hat\partial S_{0} i\hat\partial |\Phi_{0I}> = im
<\Phi_{0I}|S_{0}i\hat\partial |\Phi_{0I}>
\end{equation}
The matrix element $ <\Phi_{0I}|S_I|\Phi_{0I}>  =\frac{1}{im}, $
 more over
\begin{equation}
S_I|\Phi_{0I}> = \frac{1}{im}|\Phi_{0I}>,\,\,\, <\Phi_{0I}|S_I
=<\Phi_{0I}|\frac{1}{im}
\end{equation}
as it must be.

In the field of instanton ensemble, represented by $A=\sum_I A_I$,
full quark propagator, expanded with respect to a single
instanton, and with account Eq. \re{Si} is:
\bea
&&S =  S_{0} + \sum_I (  S_{I} -  S_{0}) + \sum_{I\neq J}
 (  S_{I} -  S_{0})  S_{0}^{-1} (  S_{J} -  S_{0})
 \nonumber\\
 && + \sum_{I\neq J,\,J\neq K}  (  S_{I} -  S_{0})
  S_{0}^{-1}
 (  S_{J} -  S_{0})  S_{0}^{-1}(  S_K-  S_{0})+...
\nonumber\\
 &&=S_{0} + \sum_{I,J} S_{0} i\hat\partial |\Phi_{0I}>
\left(\frac{1}{C}+\frac{1}{C}T\frac{1}{C}+...\right)_{IJ}
<\Phi_{0J}|i\hat\partial S_{0}
\nonumber\\
&&=S_{0} + \sum_{I,J}S_{0} i\hat\partial
|\Phi_{0I}>\left(\frac{1}{C-T}\right)_{IJ}
<\Phi_{0J}|i\hat\partial S_{0}
\label{propagator1}
\eea
where
\bea
C_{IJ}=\delta_{IJ}c_I = -\delta_{IJ} <\Phi_{0I}|i\hat\partial
S_{0} i\hat\partial|\Phi_{0I}>,
\nonumber
\\
(C-T)_{IJ} = -<\Phi_{0I}|i\hat\partial S_0 i\hat\partial
|\Phi_{0J}>
\eea
We are calculating $\Det_{low}$ using the formula:
\begin{equation}
\ln{\Det}_{low}= \Tr\int^m_{M_1}idm'(\tilde S(m')- \tilde
S_{0}(m'))
\end{equation}
Within zero-mode assumption (Eq. \re{Si}) the trace is restricted
to the subspace of instantons:
\begin{equation}
\Tr (S - S_0 )= -\sum_{I,J} <\Phi_{0,J}|i\hat\partial\, S_{0}^2\,
i\hat \partial |\Phi_{0,I}> <\Phi_{0,I}|(\frac{1}{i\hat\partial
S_{0}i\hat\partial}) |\Phi_{0,J}>
\end{equation}
Introducing now the matrix
\bea
 B(m)_{IJ}= <\Phi_{0,I}|i\hat\partial  S_0
i\hat\partial |\Phi_{0,J}>
\eea
it is easy to show that
\bea
\ln{\Det}_{low}=\Tr\int^m_{M_1}idm'( S(m')-  S_{0}(m'))
 = \sum_{I} \int^{ B(m)}_{ B (M_1)}
(d  B(m')\frac{1}{ B(m')})_{II}
\nonumber
\\
= \Tr \ln \frac{ B(m)}{ B(M_1)}=\ln\det B(m) - \ln\det B(M_1)
\lab{detlow}
\eea
which is desired answer. The determinant $\det B(m)$ from Eq.
\re{detlow} is the extension of the Lee-Bardeen result
\cite{Lee:sm} for the non-small values of current quark mass $m$.

\section*{Light quark effective action beyond chiral limit}

Averaged $\Det_{low}$ leads to the partition function $Z[m]$,
which for $N_f =2$ has the form:
\bea
Z[m] &=& \int d\lambda_+ d\lambda_- D\psi D\psi^{\dagger} \exp
[\int d^4 x \sum_{f=1}^{2}\psi_{f}^{\dagger}(i\hat\partial \,+\,
im_{f})\psi_{f}
\label{Z}
\\
&& + \lambda_+ Y_{2}^+   + \lambda_- Y_{2}^- + N_+
\ln\frac{K}{\lambda_+} + N_- \ln\frac{K}{\lambda_- }],
 \nonumber
\eea
here $\lambda_\pm$ are dynamical couplings ($K$ is unessential
constant, which provide under-logarithm expression dimensionless)
\cite{Diakonov:1995qy,Musakhanov:1998wp,Musakhanov:vu}. Values of
them are defined by saddle-point calculations. $Y_{2}^\pm$ are
t'Hooft type interaction terms \cite{Diakonov:2002fq}:
\bea
Y_{2}^\pm = \frac{1}{N_{c}^2 -1}\int d^4 x [(1-\frac{1}{2N_c})
\det\, iJ^{\pm}(\rho , x) + \frac{1}{8N_c} \det\,
iJ^{\pm}_{\mu\nu}( x)]
\\\nonumber
J^{\pm}_{fg}( x) = \int \frac{d^4 k_f d^4 l_g}{(2\pi )^8}\exp
i(k_f -l_g )x q^{+}_f (k_f )\frac{1\pm \gamma_5}{2}q_g(l_g )
\\\nonumber
J^{\pm}_{\mu\nu ,fg}( x) = \int \frac{d^4 k_f d^4 l_g}{(2\pi
)^8}\exp i(k_f -l_g )x
 q^{+}_f (k_f )\frac{1\pm \gamma_5}{2}\sigma_{\mu\nu}q_g (l_g )
\eea
 where $q (k) = 2\pi \rho F(k )\psi (k)$. The form-factor $F(k)$ is
due to zero-modes and has explicit form $ F(k) = -
\frac{d}{dt}[I_0 (t )K_0 (t ) - I_1 (t )K_1 (t )]_{t
=\frac{|k|\rho}{2}} . $ In the following we will neglect by
$J^{\pm}_{\mu\nu ,fg}(x)$ interaction term, since it give a
$O(\frac{1}{N_{c}^2})$ contribution to the quark condensate. Since
$
q(x)=\int \frac{d^4 k}{(2\pi )^4}\exp (ikx) \,\, q(k),\,\,\,\,
J^{\pm}_{fg}(x) = q^{+}_f (x)\frac{1\pm \gamma_5}{2}q_g (x ),
$
\bea
\lab{detJ}
&&\det \frac{iJ^+ ( x)}{g} + \det \frac{iJ^- ( x)}{g}
\\\nonumber
&=&\frac{1}{8g^2} ( -(q^+ (x) q (x))^2 - (q^+ (x)i\gamma_5
\vec\tau q(x))^2 + (q^+ (x)\vec\tau q(x))^2 + (q^+ (x)i\gamma_5
q(x))^2 ).
\eea
Here color factor $g^2 = \frac{(N_{c}^{2}-1)2N_c}{(2N_c -1)}.$

In the following we will take equal number of instantons and
antiinstantons $N_+=N_-=N/2$ and corresponding couplings
$\lambda_\pm=\lambda$.

 Now it is natural
to bosonize quark-quark interaction terms \re{detJ} by
introducing meson fields. For $N_f=2$ case it is exact procedure.
We have to take into account the
changes of $q$ and $q^\dagger$ under the $SU(2)$ chiral
transformations:
$$
\delta q=i\gamma_5 \vec\tau\vec\alpha q,\,\,\,\, \delta q^+ =q^+
i\gamma_5 \vec\tau\vec\alpha
$$
to introduce appropriate meson fields,
changing under $SU(2)$ chiral transformations as:
$$
 \delta\sigma =
2\vec\alpha\vec\phi,\,\,\,\,\delta\vec\phi=-2\vec\alpha \sigma
,\,\,\,\,
\delta\eta=-2\vec\alpha\vec\sigma,\,\,\,\delta\vec\sigma=2\eta\vec\alpha .
$$
Then $ \delta q^+ (\sigma+i\gamma_5\vec\tau\vec\phi )q =0,\,\,\,\,
\delta q^+ (\vec\tau\vec\sigma + i\gamma_5\eta )q =0 $ means that
these combinations of fields are chiral invariant
\footnote{Certainly, quark-quark interaction term Eq. \re{detJ} is
non-invariant over $U(1)$ axial transformations, as it must be.}.
So, the interaction term has an exact bosonized representation:
\bea
&& \int d^4 x \exp [\lambda (\det \frac{iJ^+}{g} + \det
\frac{iJ^-}{g} )]
\\ \nonumber
&& = \int D\sigma D\vec\phi D\eta D\vec\sigma
 \exp\int d^4x [ \frac{\lambda^{0.5}}{2g}
  q^+ i(\sigma +i\gamma_5 \vec\tau\vec\phi
  +i\vec\tau\vec\sigma +\gamma_5\eta)q
-\frac{1}{2}
 (\sigma^2 + {\vec\phi}^2 + {\vec\sigma}^2 +\eta^2 )]
\nonumber
\eea
Then the partition function is
\bea
&&Z[m]=\int d\lambda D\sigma D\vec\phi D\eta D\vec\sigma
\exp[N\ln\frac{K}{\lambda}-N
\\\nonumber
&&- \frac{1}{2}\int d^4 x
(\sigma^2+{\vec\phi}^2+{\vec\sigma}^2+\eta^2)+ \Tr\ln\frac{\hat p
+im+i\frac{\lambda^{0.5}}{2g}( 2\pi
\rho)^2F(\sigma+i\gamma_5\vec\tau\vec\phi+i\vec\tau\vec\sigma+\gamma_5\eta)F
}{\hat p+im }]
\eea
({$\Tr(...)$ means here $\tr_{\gamma,c,f}\int d^4x<x|(...)|x>$,
where $\tr_{\gamma,c,f}$ is the trace over Dirac, color, and
flavor indexes.}) In the following we assume $m_u = m_d = m$. Then
common saddle point on $\lambda$, $\sigma \,\,(= const)$ ($others
=0$)is defined by Eqs. $\frac{\partial V[m,\lambda , \sigma
]}{\partial \lambda}= \frac{\partial V[m,\lambda , \sigma
]}{\partial \sigma}=0,$ where the potential
\bea
V[m,\lambda , \sigma ] = -N\ln\frac{K}{\lambda}+N +
\frac{1}{2}V \sigma^2 - \Tr\ln\frac{\hat p
+i(m+M(\lambda,\sigma)F^2(p))}{\hat p+im }
\eea
and we defined $M(\lambda,\sigma)=\frac{\lambda^{0.5}}{2g}( 2\pi \rho)^2\sigma$.
Then the common saddle-point on $\lambda$ and $\sigma$ is given
by Eqs.:
\begin{equation}
N=\frac{1}{2}\Tr \frac{iM(\lambda,\sigma)F^2(p)}
{\hat p+i(m+M(\lambda,\sigma)F^2(p))}=\frac{1}{2}V\sigma^2 .
\lab{saddle}
\end{equation}
The solutions of this Eqs. are 
$\lambda_0$ and $\sigma_0=(2\frac{N}{V})^{1/2}=2^{1/2} R^{-2}$. 
It is clear that $M_0=M(\lambda_0,\sigma_0)$ 
has a meaning of dynamical quark mass, which is defined by this Eqs..
 At typical values $R^{-1}=200 \,MeV,\,\,\rho^{-1}=600\,MeV$ we have
$\sigma^2_0=2(200\,MeV)^4$, and in chiral limit $m=0$
$M_0\rightarrow M_{00}=358\,MeV$, $\lambda_{00}\approx M_{00}^2$.  
It is clear that due to
saddle-point equation \re{saddle} $M_0$ (and $\lambda_0$) become
the function of the current mass $m$. This dependence was
investigated in \cite{Musakhanov:vu}.

\section*{Vacuum with account of quantum corrections}

The account of the quantum fluctuations around saddle-points $\sigma_0,\lambda_0$ 
will change the potential $V[m,\lambda , \sigma ]$ to $V_{eff}[m,\lambda , \sigma ]$
(it is clear that the difference between these two potentials 
is order of $1/N_c$).
Then, the partition function is given by Eq.
\begin{equation}
Z[m]=\int d\lambda \exp(-V_{eff}[m,\lambda,\sigma])
\end{equation}
There is important difference between this instanton 
generated partition function $Z[m]$ and 
traditional $NJL$-type models -- we have to
integrate over the coupling $\lambda$ here.
As was mentioned before, this integration on $\lambda$
by saddle-point method leads to exact answer.
This saddle-point is defined by Eq.:
\begin{equation}
\frac{dV_{eff}[m,\lambda,\sigma]}{d\lambda}=0
\label{saddlelambda}
\end{equation}
which leads to the $\lambda$ as a function of $\sigma$,
i.e. $\lambda=\lambda(\sigma)$.

Then, the vacuum is the minimum of the effective potential
 $V_{eff}[m,\sigma]$, which is given by a solution of the equation
\begin{equation}
\frac{d V_{eff}[m,\sigma,\lambda(\sigma)]}{d\sigma}=
\frac{\partial V_{eff}[m,\sigma,\lambda(\sigma)]}{\partial\sigma}=0.
\lab{vacuum}
\end{equation}
where it was used  Eq. \re{saddlelambda}.
 
  We denote a fluctuations as a primed fields $\Phi'_i$.
 The action and corresponding $V_{eff}$ now has a form:
\bea
&& S[m,\lambda,\sigma,\Phi']=S_0[m,\lambda,\sigma]+S_V[m,\lambda,\sigma,\Phi'],
\\\nonumber
&&S_0[m,\lambda,\sigma]=V[m,\lambda,\sigma]= \frac{1}{2}V\sigma^2
-\Tr\ln\frac{\hat p +i(m+M(\lambda,\sigma)F^2)}{\hat p+im } -N\ln\frac{K}{\lambda}+N
\\\nonumber
&& S_V[m,\lambda,\sigma,\Phi']= \int
d^4x\frac{1}{2}({\sigma}^{'2}+{\vec\phi}^{'2}+{\vec\sigma}^{'2}+{\eta}^{'2})
\\\nonumber
&& -\frac{1}{2\sigma^2}\Tr\left[\frac{iM(\lambda,\sigma)F^2}
{\hat p+i(m+M(\lambda,\sigma)F^2)} (\sigma'+i\gamma_5\vec\tau\vec\phi'
+i\vec\tau\vec\sigma'+\gamma_5\eta')\right]^2 ,
\eea
and
\bea
V_{eff}[m,\lambda,\sigma]= S_0[m,\lambda,\sigma]+
 V_{eff}^{mes}[m,\lambda,\sigma]
\lab{Veff}
\eea
Here second term in Eq. \re{Veff} is explicitly represented by
 \bea
\lab{loops}
&&V_{eff}^{mes}[m,\lambda,\sigma]=\frac{1}{2}\Tr\ln\frac{\delta^2
 S_V[m,\lambda,\sigma,\Phi']}{\delta\Phi_i'(x)\delta\Phi_j'(y)}
 =\frac{V}{2}\sum_i\int\frac{d^4q}{(2\pi)^4}
 \ln [1-\tr\frac{1}{\sigma^2}\int\frac{d^4p}{(2\pi)^4}
\nonumber\\
&&\times \frac{M(\lambda,\sigma) F^2(p)}{\hat
p+i(m+M(\lambda,\sigma)F^2(p))}\Gamma_i\frac{M(\lambda,\sigma) F^2(p+q)}{\hat p+\hat q
+i(m+M(\lambda,\sigma)F^2(p+q) )}\Gamma_i ],
\eea
where the factors $\Gamma_i=(1,i\gamma_5\vec\tau, i\vec\tau
,\gamma_5)$ and the sum on $i$ is counted  all corresponding meson
fluctuations $\sigma',\vec\phi',\vec\sigma',\eta'$. $\tr$ here
means the trace over flavor, color and Dirac indexes. Integrals
in Eq. \re{loops} are completely convergent one due to the
presence of the form-factors $F$.

Certainly the quantum fluctuations contribution will move the
the coupling $\lambda$ from $\lambda_0$ to $\lambda_0+\lambda_1$
and $\sigma$ as $\sigma_0\rightarrow\sigma_0
+ \sigma_1$, where $\frac{\lambda_1}{\lambda_0}$ and $\frac{\sigma_1}{\sigma_0}$ are 
of order $1/N_c$. 

First, consider Eq. \re{saddlelambda}:
\bea
\label{dlambda}
&&\lambda\frac{dV_{eff}[m,\lambda,\sigma]}{d\lambda}=
N-\frac{1}{2}\Tr\frac{iM(\lambda,\sigma)F^2}
{\hat p+i(m+M(\lambda,\sigma)F^2)}
+\frac{V}{2}\sum_i\int\frac{d^4q}{(2\pi)^4}
\\\nonumber
&&\times [\sigma^2-\tr\int\frac{d^4p}{(2\pi)^4}
\frac{M(\lambda,\sigma) F^2(p)}{\hat
p+i(m+M(\lambda,\sigma)F^2(p))}\Gamma_i\frac{M(\lambda,\sigma) F^2(p+q)}{\hat p+\hat q
+i(m+M(\lambda,\sigma)F^2(p+q) )}\Gamma_i ]^{-1}
\\\nonumber
&&\times[-\tr\int\frac{d^4p}{(2\pi)^4}
\frac{M(\lambda,\sigma) F^2(p)}{\hat
p+i(m+M(\lambda,\sigma)F^2(p))}\Gamma_i\frac{M(\lambda,\sigma) F^2(p+q)}{\hat p+\hat q
+i(m+M(\lambda,\sigma)F^2(p+q) )}\Gamma_i 
\\\nonumber
&&+ i\tr\int\frac{d^4p}{(2\pi)^4}
\left(\frac{M(\lambda,\sigma) F^2(p)}{\hat
p+i(m+M(\lambda,\sigma)F^2(p))}\right)^2
\Gamma_i\frac{M(\lambda,\sigma) F^2(p+q)}{\hat p+\hat q
+i(m+M(\lambda,\sigma)F^2(p+q) )}\Gamma_i ]=0
\eea
From this saddle-point Eq. we get $\lambda=\lambda(\sigma)$.

From vacuum Eq. \re{vacuum} we in similar manner arrive to: 
\bea
\label{dsigma}
&&\sigma\frac{\partial V_{eff}[m,\sigma,\lambda(\sigma)]}{\partial\sigma}=
V\sigma^2 -\Tr\frac{iM(\lambda(\sigma),\sigma)F^2}
{\hat p+i(m+M(\lambda(\sigma),\sigma)F^2)}
+\frac{V}{2}\sum_i\int\frac{d^4q}{(2\pi)^4}
\\\nonumber
&&\times [\sigma^2-\tr\int\frac{d^4p}{(2\pi)^4}
\frac{M(\lambda(\sigma),\sigma) F^2(p)}{\hat
p+i(m+M(\lambda(\sigma),\sigma)F^2(p))}\Gamma_i\frac{M(\lambda(\sigma),\sigma) 
F^2(p+q)}{\hat p+\hat q
+i(m+M(\lambda(\sigma),\sigma)F^2(p+q) )}\Gamma_i ]^{-1}
\\\nonumber
&&\times[ 2 i\tr\int\frac{d^4p}{(2\pi)^4}
\left(\frac{M(\lambda(\sigma),\sigma) F^2(p)}{\hat
p+i(m+M(\lambda(\sigma),\sigma)F^2(p))}\right)^2
\Gamma_i\frac{M(\lambda(\sigma),\sigma) F^2(p+q)}{\hat p+\hat q
+i(m+M(\lambda(\sigma),\sigma)F^2(p+q) )}\Gamma_i ]=0
\eea
Since we are believing to $\frac{1}{N_c}$ expansion, it is natural 
inside quantum fluctuations contribution (under the integrals over
$q$) to take $\sigma=\sigma_0$, $M(\lambda(\sigma),\sigma)=M_0$. 

To simplify the expressions
introduce  vertices $V_{2i}(q)$,  $V_{3i}(q)$ 
and meson propagators $\Pi_i (q)$, which  are defined as:
\bea
\lab{vertices}
&& V_{2i}(q) =\tr\int\frac{d^4p}{(2\pi)^4}
 \frac{M_0(p)}{\hat p+i\mu_0(p)}\Gamma_i
 \frac{M_0(p+q)}{\hat p+\hat q+i\mu_0(p+q)}\Gamma_i
 \\
&& V_{3i}(q) =\tr\int\frac{d^4p}{(2\pi)^4}
 \left(\frac{M_0(p)}{\hat p+i\mu_0(p)}\right)^2\Gamma_i
 \frac{M_0(p+q)}{\hat p+\hat q+i\mu_0(p+q)}\Gamma_i
 \\
 \lab{propagators}
 &&\Pi^{-1}_i (q)=\frac{2}{R^4} - V_{2i}(q).
\eea
Here $M_0(p)=M_0F^2(p),\,\,\,\mu_0(p)=m+M_0(p)$ 
and was taken into account that
$\sigma_0^2=2R^{-4}$.

From Eqs. \re{dlambda} and \re{dsigma} we have 
\bea
&&\frac{M_1}{M_0}\left[\frac{2}{R^4}+\frac{1}{V}\Tr\left(\frac{M_0(p)}
{\hat p+i\mu_0(p)}\right)^2\right]
=\sum_i\int\frac{d^4q}{(2\pi)^4} (iV^3_i(q)-V_{2i}(q))\Pi_i (q)
\label{M1}
\\
&&\frac{\sigma_1}{\sigma_0} 
=-\frac{R^4}{4}\sum_i\int\frac{d^4q}{(2\pi)^4} V_{2i}(q)\Pi_i (q)
\label{sigma2}
\eea
The vertices $V_{2i}(q)$, $V_{3i}(q)$ and the meson
propagators $\Pi_i(q)$ are well defined functions, providing
well convergence of the integrals in Eqs. \re{M1}, \re{sigma2}.

It is of special attention to {\bf the contribution of pion
fluctuations  $\vec\phi'$ at small pion momentum $q$. } We shall
demonstrate that this contribution leads to the famous chiral log
term with model independent coefficient in the correspondence
with previous calculations in NJL-model \cite{Nikolov:1996jj}.

Pion inverse propagator of $\Pi^{-1}_{\vec\phi'}(q)$ at small
$q\sim m_\pi$ is:
$
\Pi^{-1}_{\vec\phi'}(q) =  f_{kin}^2(m_{\pi}^2 + q^2).
$
At lowest order on $m$, $f_{kin,m=0}\approx f_\pi = 93\,MeV$,
$m_\pi^2\sim m$.

The vertices in the right side of  Eq. \re{M1}
at $q=0$ and in chiral limit are:
\bea
V_{2\vec\phi_i', m=0}(0)=\frac{2}{R^4},\,\,\,\,i V_{3\vec\phi_i', m=0}(0) - V_{2\vec\phi_i', m=0}(0)
 = 8N_c\int\frac{d^4p}{(2\pi)^4}\frac{p^2M_0^2(p)}{(p^2+M_0^2(p))^2}
\eea
We see that the factor in the left side of Eq. \re{M1}
 in the chiral limit is equal to:
\bea
\tr\int\frac{d^4p}{(2\pi)^4}\frac{i\hat p M_0(p)}{(\hat p+iM_0(p))^2}=
-2(i V_{3\vec\phi_i', m=0}(0) - V_{2\vec\phi_i', m=0}(0)) 
\eea
Collecting all the factors we get small $q\leq \kappa$
contribution of pion fluctuations  $\vec\phi'$:
\bea
\lab{M1-phi-smallq}
\frac{\sigma_1}{\sigma_0}|_{\vec\phi', small\,
q}&=&\frac{M_1}{M_0}|_{\vec\phi', small\,
q} =-\frac{3}{2f_\pi^2}\int_0^{\kappa} \frac{d^4q}{(2\pi)^4}
\frac{1}{m_{\pi}^2+ q^2}
\\\nonumber
&=&-\frac{3}{32\pi^2f_\pi^2}\int_0^{\kappa^2} q^2
dq^2\frac{1}{f_{\pi}^2(m_{\pi}^2+
q^2)}=-\frac{3}{32\pi^2f_\pi^2}(\kappa^2
+m_\pi^2\ln\frac{m_\pi^2}{\kappa^2+m_\pi^2})
\eea
Here we put $m=0$ everywhere except $m_\pi$. We see that the
coefficient in the front of of $m_\pi^2\ln m_\pi^2$ is a model
independent as it must be.

Quick estimate, assuming $\kappa=\rho^{-1}$, gives  
\bea
\lab{M1-phi}
\frac{\sigma_1}{\sigma_0}|_{\vec\phi'}\approx\frac{M_1}{M_0}|_{\vec\phi'}
\approx-\frac{3}{32\pi^2f_\pi^2\rho^2}(1
+m_\pi^2\rho^2\ln m_\pi^2\rho^2)\approx -0.4 (1+0.054\ln 0.054)
\eea
So, we expect that pion loops is provided not only non-analytical 
$\frac{1}{N_c}m\ln m$ term but also very 
large contribution to $\frac{1}{N_c}$ term. 
 
This one dictate the strategy of the following calculations of $\sigma_1$ and $M_1$:\\
1. we have to extract analytically $\frac{1}{N_c}m\ln m$ term from pion loops;\\
2. rest part of $\sigma_1$ and $M_1$ can be calculated numerically and expanded
over $m$, paying special attention to the pion loops
and keeping $\frac{1}{N_c}$ and $\frac{1}{N_c}m$ terms.

For actual numerical calculations we are using simplified version
of the form-factor $F(p)$ from \cite{Petrov:1998kg} (with
corrected high momentum dependence):
\bea
\lab{F(p)}
F(p<2 GeV)=\frac{L^2}{L^2+p^2},\,\,\, F(p>2 GeV)=\frac{1.414}{p^3}
\eea
 where $L \approx \frac{\sqrt{2}}{\bar\rho}= 848 MeV$.

 At $N_c=3$ semi-numerical calculations of $M_1$ and $\sigma_1$   
 lead to:
\bea
&&\frac{M_1}{M_0}
=-0.662 -4.64 m - 4.01 m\ln m
\lab{formulaM1}
\\
&&\frac{\sigma_1}{\sigma_0}
=-0.523 - 4.26\, m - 4.00\, m \ln m
\lab{formula}
\eea
Here $m$ is given in $GeV$. Certainly, in \re{formulaM1}  the
$m\,\ln m$ term is completely correspond to Eq.
\re{M1-phi-smallq}. $\frac{M_1}{M_0}$ is $-66\%$ in chiral limit and
reach its maximum $\sim -20\%$ at $m\sim 0.115\, GeV.$

The relative shift of the
vacuum 
$\frac{\sigma_1}{\sigma_0} $ is $-52\%$  at the chiral limit 
and reach its maximum
$\sim -2\%$ at $m\sim 0.125 \, GeV.$

The main contribution to both quantities $\frac{M_1}{M_0}$ and  
$\frac{\sigma_1}{\sigma_0}$ 
come from pion loops. Other mesons  give the contribution $\sim 10\%$ to 
$O(\frac{1}{N_c})$ and $O(\frac{1}{N_c}m)$ terms.

\section*{Quark condensate}

We have to calculate quark condensate beyond chiral limit taking
into account $O(m),$ $O(\frac{1}{N_c})$, $O(\frac{1}{N_c} m)$ and
$O(\frac{1}{N_c}m\ln m)$ terms. Quark condensate is extracted from
the partition function:
\bea
\nonumber
<\bar qq>&=&\frac{1}{2V}\frac{d V_{eff}[m,\lambda,\sigma]}{d m}
=\frac{1}{2V}\frac{\partial
(V[m,\lambda,\sigma]+V^{mes}_{eff}[m,\lambda_0,\sigma_0])} {\partial m}
\\
&=&-\frac{1}{2V}\Tr(\frac{i}{\hat p +i\mu (p)}-\frac{i}{\hat
p+im}) +\frac{1}{2V}\frac{\partial V^{mes}_{eff}[m,\lambda_0,\sigma_0]}
{\partial m}
\lab{cond}
\eea
here $\lambda=\lambda_0+\lambda_1,\,\,\,\,, \sigma=\sigma_0+\sigma_1,\,\,\,\
M=M_0+M_1,\,\,\,\mu(p)=m+MF^2(p)$. First term of Eq. \re{cond}
is
\bea
&&-\frac{1}{2V}\Tr(\frac{i}{\hat p +i\mu(p)}-\frac{i}{\hat
p+im})
\\\nonumber
&&=-4N_c\int\frac{d^4p}{(2\pi)^4}\left(\frac{\mu_{0}(p)}{p^2+\mu_{0}^2(p)}
-\frac{m}{p^2+m^2}
+\frac{M_1}{M_0}\frac{M_0(p)(p^2-\mu_{0}^2(p))}{(p^2+\mu_{0}^2(p))^2}\right)
\eea
Second term of Eq. \re{cond} -- meson loops contribution to the
condensate is
\bea
\nonumber
\frac{1}{2V}\frac{\partial V_{eff}^{mes}[m,\lambda_0,\sigma_0]}
{\partial m} &=& \frac{i}{2}\sum_i\int\frac{d^4q}{(2\pi)^4}
 \left(\tr\int\frac{d^4p}{(2\pi)^4}
 \frac{M_0(p)}{(\hat p+i\mu_0(p))^2}\Gamma_i
 \frac{M_0(p+q)}{\hat p+\hat q+i\mu_0(p+q)}\Gamma_i\right)
\\
 &&\times\left(\frac{2N}{V} -
 \tr\int\frac{d^4p}{(2\pi)^4}\frac{M_0(p)}{\hat p+i\mu_0(p)}
  \Gamma_i \frac{M_0(p+q)}{\hat p+\hat
  q+i\mu_0(p+q)}\Gamma_i\right)^{-1}
\eea
At $m=0$ and without meson loops the condensate is
\bea
<\bar qq>_{00}=
-4N_c\int\frac{d^4p}{(2\pi)^4}\frac{M_{00}(p)}{p^2+M_{00}^2(p)}
\eea
Here $M_{00}\equiv M_{0,m=0}$.

Let us to consider now the contribution of pion fluctuations
$\vec\phi'$ to the quark condensate at small $q$. First we
consider:
\bea
 \frac{1}{2V}\frac{\partial V_{eff}^{\vec\phi',\, small\, q
}[m,\lambda_0,\sigma_0]}{\partial m}= 12N_c\int\frac{d^4
p}{(2\pi)^4}\frac{M_0^2(p)\mu_0(p)}{(p^2+\mu_0^2(p))^2}
\int_0^\kappa \frac{d^4 q}{(2\pi)^4 f_{kin}^2 (m_{\pi}^2+ q^2)}
\eea
We keep $m$ only in $m_\pi^2$. Then at $m=0$ $\mu_0(p)\Rightarrow
M_0(p)\Rightarrow M_{00}(p)$, $f_{kin}\Rightarrow f_\pi$ and we
have
\bea
<\bar qq>&=&<\bar qq>_{00} -\frac{M_1}{M_0}
4N_c\int\frac{d^4p}{(2\pi)^4}\frac{M_{00}(p)(p^2-M_{00}^2(p))}
{(p^2+M_{00}^2(p))^2}
\\\nonumber
&+&12N_c\int\frac{d^4p}{(2\pi)^4}\frac{M_{00}^3(p)}{(p^2+M_{00}^2(p))^2}
\int_0^\kappa \frac{d^4 q}{(2\pi)^4}\frac{1}{ f_{\pi}^2
(m_{\pi}^2+ q^2)}
\\\lab{cond-phi-small-q}
&&=<\bar qq>_{00}\left(1-\frac{3}{2}\int_0^\kappa \frac{d^4
q}{(2\pi)^4}\frac{1}{ f_{\pi}^2 (m_{\pi}^2+ q^2)} \right)
\eea
Eq. \re{M1-phi-smallq} for $\frac{M_1}{M_0}$ was applied here. We
see that Eq. \re{cond-phi-small-q} is in the full correspondence
with \cite{Gasser:1983yg,Novikov:xj}.

Detailed numerical calculations lead to the semi-analytical
formula for the quark condensate including
 all $O(m),$ $O(\frac{1}{N_c})$ and
$O(\frac{1}{N_c}m\ln m)$-corrections:
\begin{equation}
<\bar qq>=<\bar qq>_{m=0}\left(1 - 18.53\, m - 7.72\, m \ln m\right)
\label{cond1}
\end{equation}
Here $<\bar qq>_{m=0}= 0.52 <\bar qq>_{00}$.
Certainly, the $ m \ln m$ term in Eq.\re{cond1} is
in full correspondence with  Eq. \re{cond-phi-small-q}, as it must be.
$<\bar qq>/<\bar qq>_{m=0}$ is a rising 
function of $m$ until $m\sim 0.04 \,GeV$ and is a falling one in the region
 $m > 0.04 \,GeV.$ 

The main contribution  to $O(\frac{1}{N_c})$, $O(\frac{1}{N_c}m)$ and
$O(\frac{1}{N_c}m\ln m)$ terms in $\frac{<\bar qq>}{<\bar qq>_{00}}$
is due to pion loops. Other mesons
give the contribution $\sim few\,\%$ to $O(\frac{1}{N_c})$ 
and $O(\frac{1}{N_c}m)$ terms.

\section*{ $m_d-m_u$ effects in quark condensate}

Current quark mass become diagonal $2\times 2$ matrix with
$m_1=m_u, m_2=m_d$,  $m=
m_1\frac{1+\tau_3}{2}+m_2\frac{1-\tau_3}{2}=m+\delta
m\frac{\tau_3}{2}$. Here $m=\frac{m_1+m_2}{2}, \delta m= m_1-m_2$.
 Let us  introduce external
field $s_i$. In our particular case it is $s_3=i\frac{\delta
m}{2},s_1=s_2=0 $. Our aim is to find the asymmetry of the quark 
condensate $\frac{<\bar uu>-<\bar dd>}{<\bar uu>}$, 
taking into account only $O(\delta m)$ terms and neglecting by
$O(\frac{1}{N_c}\delta m),$ $O(\frac{1}{N_c}\delta m\ln m).$  
It means that we  neglect at all 
by meson loops contribution.

In the presence of the external field $\vec s$ we expect also vacuum
field $\vec\sigma$. Effective potential  within requested accuracy is
\bea
V_{eff}[\sigma,\vec\sigma, m]&\approx& S_0[m,\lambda,\sigma,\vec\sigma]
\\\nonumber
&&= \frac{V}{2}(\sigma^2 +{\vec\sigma}^2)
-\Tr\ln\frac{\hat p + i\vec\tau\vec
s +i(m+M(\lambda,\sigma,\vec\sigma)F^2)}{\hat p+im+ i\vec\tau\vec
s } -N\ln\frac{K}{\lambda}+N.
\eea
$\lambda, \sigma, \vec \sigma$ are defined by the vacuum equations:
\begin{equation}
\frac{\partial V_{eff}}{\partial\lambda}=0,\,\,\,
\frac{\partial V_{eff}}{\partial\sigma}=0,\,\,\,\,
\frac{\partial V_{eff}}{\partial\sigma_i}=0.
\end{equation}
 They can be reduced to the following form:
\bea
\frac{1}{2}\Tr \frac{F^2(p)M_i(m_i+M_iF^2(p)) }{p^2+ (m_i
+M_iF^2(p))^2}=N
\eea
where 
$M_i=\frac{\lambda^{0.5}}{2g}( 2\pi \rho)^2(\sigma \pm \sigma_3)$.
Solution of these equations leads to $\lambda=\lambda [m,\vec s]$,
$\sigma =\sigma[m,\vec s]$ $\sigma_i=\sigma_i[m,\vec s]$. We have to put
them into $V_{eff}$ and find $V_{eff}=V_{eff}[m,\vec s]$. Desired
correlator is
\begin{equation}
\frac{\partial V_{eff}[m,\vec s]}{\partial s_3}|_{s_3=\frac{\delta m}{2},s_{1,2}=0}
\end{equation}
We calculate this correlator within requested accuracy, 
taking into account only  $O(\delta m)$ terms. 
So, the difference of the vacuum quark condensates of $u$ and $d$
quarks is
\bea
&&<\bar uu>-<\bar dd>
\\\nonumber
&&=\frac{1}{V}\left[ \Tr(\frac{-i}{\hat p+i(m_u+M_uF^2)}-\frac{-i}{\hat p+im_u})
-\Tr(\frac{-i}{\hat p+i(m_d+M_dF^2)}-\frac{-i}{\hat p+im_d})\right]
\eea
We expect that $ <\bar dd>\, <\, <\bar
uu>$ if $m_d\,>\,m_u$.

Typical values of light current quark masses
\cite{Leutwyler:1996eq} are $m_u =5.1 MeV$, $m_d=9.3 MeV$ on the
scale $1GeV$ (which is in fact close to our scale
$\rho^{-1}=0.6\, GeV $) leads to the asymmetry
\begin{equation}
\frac{<\bar uu>-<\bar dd>}{<\bar uu>}= 0.026
\lab{asym}
\end{equation}
From this asymmetry and using sum-rules \cite{Gasser:1984gg} we
estimate strange quark condensate at $m_s=120 \, MeV$ as:
\begin{equation}
\frac{<\bar ss>}{<\bar uu>}= 0.43,
\end{equation}
which is rather small. The reason that the asymmetry \re{asym} is
rather large.

\section*{Conclusion}

In the framework of instanton vacuum model it was calculated
simplest possible correlator -- quark condensate with complete
account of $O(m)$, $O(\frac{1}{N_c})$, $O(\frac{1}{N_c}m)$  
and $O(\frac{1}{N_c}m\ln m)$ terms,
demanding the calculation of meson loops contribution. Since initial
instanton generated  quark-quark interactions are nonlocal and
contain corresponding form-factor induced by quark zero-mode,  these loops
correspond completely convergent integrals. 
The main loop corrections come from the pions, as it was expected. 
We found that  $O(\frac{1}{N_c})$ corrections are very large $\sim 50\%$, which
request the $\sim 10\%$ changing of the basic parameters -- 
average inter-instanton distance $R$ and average instanton size $\rho$ to restore
chiral limit value of the quark condensate $<\bar qq>_{m=0}$ 
and other important quantities as $f_\pi$ and $m_\pi$ 
to their phenomenological values. This work in the progress.

In general, it was demonstrated, that instanton
vacuum model is well working tool also beyond chiral limit and
satisfy chiral perturbation theory.

\section*{Acknowledgement}
I am very grateful
to D. Diakonov for a useful discussions.
This work was done in the collaboration with Hyun-Chul Kim and M. Siddikov
and was partly supported by Republic of Korea "Brain Pool" Program.

\end{document}